\documentclass[12pt]{iopart}

\usepackage{graphicx}
\usepackage{dcolumn}
\usepackage{float}
\usepackage{iopams}

\begin{document}

\title{\boldmath Manifestation of pseudogap in ab-plane optical characteristics \unboldmath}

\author{J. Hwang$^{1}$, J. Yang$^{1}$, J. P. Carbotte$^{1,2}$ and T. Timusk$^{1,2}$}
\address{$^{1}$Department of Physics and Astronomy, McMaster University,
Hamilton, Ontario L8S 4M1, Canada\\ $^{2}$The Canadian Institute
for Advanced Research, Toronto, Ontario M5G 1Z8, Canada}

\ead{jhwang@phys.ufl.edu}

\date{\today}

\begin{abstract}
A model in which a gap forms in the renormalized electronic
density of state (DOS) with missing states recovered just above
the pseudogap $\Delta_{pg}$, is able to give a robust description
of the striking, triangular like, peak seen in the real part of
the optical self-energy of underdoped cuprates. We use this model
to explore the effect of the pseudogap on the real part of the
optical conductivity and on the partial sum rule associated with
it. An important result is that the optical spectral weight
redistributes over a much larger frequency window than it does in
the DOS.
\end{abstract}


\maketitle

\section{Introduction}

Cuprate superconductors undergo considerable change in their
electronic structure as a function of doping. In the overdoped
regime there is evidence that Fermi liquid theory applies.
However, as the doping is reduced through the optimum and then
underdoped regime, a non-Fermi liquid state evolves which is
characterized by the formation of a
pseudopgap\cite{timusk99,zasadzinski01,sutherland03,emery95,krasnov00,
letacon06,kanigal06,warren89,homes93,loram98,loram01,renner98}.
There is as yet no consensus as to the correct microscopic
understanding of the pseudogap state. There is evidence that it
corresponds to the formation of preformed Cooper
pairs\cite{emery95} at some new characteristic temperature $T^{*}$
above the superconducting $T_c$. Superconductivity sets in at the
temperature where phase coherence among the preexisting pairs
occurs. Only then does the system acquire long range order.
Another prominent possibility for the pseudogap is competing
order\cite{chakravarty01,dora02,chakravarty01a,zhu01,yang02,wang01,chakravarty02,
capelutti99,benfatto00,aristov04,valenzuela05,kim02,kim02a,benfatto06,aristov05,gerami06}
such as d-density wave formation. On the other hand, from a
phenomenological point of view many experimental findings, in the
pseudogap region of the cuprate phase diagram, can be understood
in a model which corresponds to a reduction of the electronic
density of state in the vicinity of the Fermi energy over an
energy region ($\Delta_{pg}$) which sets the pseudogap scale. An
important concept in such a characterization of the pseudogap is
its variation with temperature. At zero temperature there is a
full gap of order $\Delta_{pg}$ which fills in with increasing
temperature but does not change its
magnitude\cite{kanigal06,renner98}. The pseudogap temperature
($T^{*}$) corresponds to compete filling rather than closing {\it
i.e.} $\Delta_{pg} \rightarrow$ 0.

Recently more details about the temperature evolution of the
pseudogap have emerged from angle-resolved photoemission (ARPES)
experiments which have been interpreted in term of Fermi
arcs\cite{kanigal06}. Below $T^{*}$ a full pseudogap opens up, but
only on a small region of the Fermi surface near the antinodal
direction. The rest of the Fermi surface, called the Fermi arc,
remains ungaped. Experiments have shown that the length of the
Fermi arc centered in the nodal direction is proportional to the
reduced temperature $t = T/T^{*}$ and vanishes at $T = 0$ at which
point the entire Fermi surface is fully gaped. The existence of a
full gap in the electronic density of state at $T = 0$ is
consistent with many other experiments in particular specific
heat. It also has implications for the behavior of the in-plane
optical conductivity.

It has recently been pointed\cite{hwang07} out that the sharp
triangular like cap observed in the real part of the optical
self-energy seen in underdoped samples of
Bi$_{2}$Sr$_{2}$CaCu$_{2}$O$_{8+\delta}$
(Bi-2212)\cite{hwang04,hwang07a} and orthoII
YBa$_{2}$Cu$_{3}$O$_{6.5}$ (YBCO$_{6.5}$)\cite{hwang06} follows
directly from the opening of a full pseudogap with the lost
electronic density of states below $\Delta_{pg}$ recovered in the
region just above it. This represents a clear signature of
pseudogap behavior in optical spectroscopy. In this paper we
consider the implication of such a phenomenological model on
optical properties.

The paper is structured as follows. In section II we introduce the
generalized Drude model for the optical conductivity which relates
it to an optical self-energy, $\Sigma^{op}(\omega)$. We also
summarize the data for the real part of $\Sigma^{op}(\omega)$ on
which our pseudogap model is based and the distinct difference
between underdoped and overdoped cases is noted and emphasized.
Our theoretical model is introduced and compared with the data. In
section III we deal with signatures of the pseudogap in the real
part of the conductivity and partial sum rule. In section IV we
conclude the paper with a summary of our findings.

\section{Theoretical model}

The optical conductivity $\sigma(\omega)$ in a correlated electron
system can be analyzed in terms of a generalized Drude form
written as\cite{hwang07,hwang04}
\begin{equation}
\sigma(T, \omega)=
\frac{i}{4\pi}\frac{\Omega_{p}^2}{\omega-2\Sigma^{op}(T, \omega)},
 \label{eq1}
\end{equation}
where $T$ is temperature, $\Omega_p$ is the plasma frequency and
$\Sigma^{op}(T, \omega)\equiv\Sigma_1^{op}(T,
\omega)+i\Sigma_2^{op}(T,\omega)$ is the optical self-energy. The
imaginary part of -2$\Sigma^{op}(T, \omega)$ is equal to the
optical scattering rate $1/\tau^{op}(T, \omega)$ and the real part
can be written in terms of an optical effective mass $m^{*, op}(T,
\omega)/m$ with $\omega[m^{*, op}(T, \omega)/m - 1] \equiv
-2\Sigma^{op}_{1}(T, \omega)$. While the optical scattering rate
and the mass renormalization $\lambda^{op}(T, \omega)$
($1+\lambda^{op}(T, \omega)\equiv m^{*, op}(T, \omega)/m$) defined
here are not the same as those defined from the quasiparticle
self-energy $\Sigma^{qp}(T, \omega)$ they are related through the
equation for the conductivity. Neglecting vertex corrections and
taking zero temperature ($T = 0$) for an isotropic system we
have\cite{marsiglio03}
\begin{equation}
\sigma(\omega)=\frac{\Omega_p^2}{4
\pi}\frac{i}{\omega}\int^{\omega}_{0}d\omega'\frac{1}{\omega+i/\tau_{imp}
-\Sigma^{qp}(\omega')-\Sigma^{qp}(\omega-\omega')} \label{eq2}
\end{equation}
where we have also included the possibility of elastic impurity
scattering through the constant scattering rate $1/\tau_{imp}$.
The quasiparticle scattering rate is $-2\Sigma^{qp}_{2}(\omega)$
and $\omega[m^{*, qp}(\omega)/m
-1]\equiv-2\Sigma^{qp}_{1}(\omega)$ in complete analogy with the
optical case. If a boson exchange theory is used to describe the
interactions among electrons, the quasiparticle self-energy at $T
= 0$ is related to the electron-boson spectral density
$I^2\chi(\omega)$ through the equation\cite{carbotte05,mitovic83}
\begin{equation}
\Sigma^{qp}(\omega)=\int^{\infty}_{0}d\Omega
I^2\chi(\Omega)\ln\Big{|}\frac{\Omega-\omega}{\Omega+\omega}\Big{|}
-i \pi \int^{|\omega|}_{0}d\Omega I^2\chi(\Omega) \label{eq3}
\end{equation}
Here we need to generalize the formalism just given to include the
possibility of an energy dependent renormalized electron density
of state (DOS) $N(\omega)$ which is defined
as\cite{mitovic83,mitrovic85,knigavko05,knigavko06,sharapov05}
\begin{equation}
N(\omega) = \sum_{\underline{k}}\frac{-Im
G(\underline{k},\omega)}{\pi} \label{eq4}
\end{equation}
where $G(\underline{k},\omega)$ is the fully renormalized Green's
function. In this case  Eq. \ref{eq3} needs modification and
reads\cite{carbotte05,mitovic83,mitrovic85,knigavko05,knigavko06,sharapov05}
\begin{equation}
\Sigma^{qp}(\omega)= 2\omega
P\int^{\infty}_{0}d\omega'\tilde{N}(\omega')\int^{\infty}_{0}
d\Omega\frac{I^2\chi(\Omega)}{\omega^2-(\omega'+\Omega)^2} - i \pi
\int^{\omega}_{0}d\Omega I^2\chi(\Omega)\tilde{N}(\omega-\Omega)
\label{eq5}
\end{equation}
Here $\tilde{N}(\omega)\equiv[N(\omega)+N(-\omega)]/2$ is the
symmetrized DOS. For the case with an energy dependent DOS the
relationship between $\Sigma^{qp}$ and $\Sigma^{op}$ is more
complicated than Eq. \ref{eq2}. There exists however simplified
equation for $\lambda^{op}(\omega)$ and $1/\tau^{op}(\omega)$
which, while not exact, are sufficiently accurate for the present
discussion. They are\cite{mitovic83,mitrovic85,knigavko05}
\begin{equation}
\lambda^{op}(\omega) = \frac{2}{\omega^2}\int^{\infty}_{0}d\Omega
I^2\chi(\Omega)P
\int^{\infty}_{0}d\omega'\tilde{N}(\omega')\ln\Big{[}
\frac{(\omega'+\Omega)^2}{(\omega'+\Omega)^2-\omega^2}\Big{]}
\label{eq6}
\end{equation}
and
\begin{equation}
\frac{1}{\tau^{op}(\omega)}=\frac{2
\pi}{\omega}\int^{\infty}_{0}d\Omega
I^2\chi(\Omega)\int^{\omega-\Omega}_{0}d\omega'\tilde{N}(\omega')
\label{eq7}
\end{equation}
%
%
\begin{figure}[t]
  \vspace*{-2.0 cm}%
  \centerline{\includegraphics[width=3.5 in]{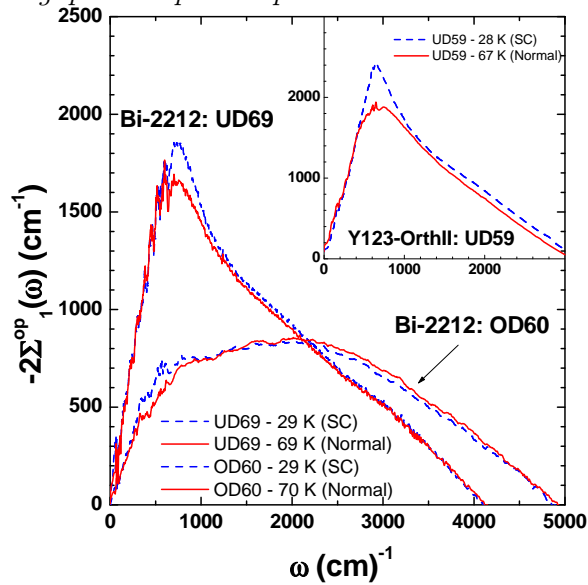}}%
  \vspace*{-2.0 cm}%
\caption{(color online) Minus twice the real part of the optical
self-energy $-2\Sigma^{op}_1(\omega)$ in units of cm$^{-1}$ as a
function of $\omega$ also in cm$^{-1}$ for an underdoped Bi-2212
with $T_c =$ 69 K (UD69) and an overdoped Bi-2212, $T_c =$ 60 K
(OD60) at two temperatures; dashed (blue) in superconducting state
and sold (red) in normal state. Shown in the inset are equivalent
results for orthoII YBCO$_{6.50}$ which is underdoped with $T_c =$
59 K (UD59).}
 \label{Fig1}
\end{figure}
In this approximation, optical and quasiparticle quantities are
related by $\Sigma^{qp}(\omega)= d/d\omega[\omega
\Sigma^{op}(\omega)]$ which can be verified through direct
differentiation of Eq. \ref{eq6} and \ref{eq7} and comparison with
Eq. \ref{eq5}. This relationship has been used in a recent
comparison of high energy scales seen in optical data with those
seen in ARPES\cite{hwang07b} which measures directly the
quasiparticle self-energy.

The exact microscopic origin of the pseudogap is not known. Here
we model it as a gap in the fully renormalized electronic density
of state $\tilde{N}(\omega)$ of formula Eq. \ref{eq4}. Such a
model has been used previously to analyze the specific
heat\cite{loram98,loram01} in the underdoped regime of the high
$T_c$ cuprates and more recently applied to optics\cite{hwang07}.
The motivating data is reproduced in Fig. \ref{Fig1} for the
convenience of the reader. These sets of data, for the real part
of the optical self-energy are presented for an in-plane
underdoped sample of Bi-2212 with a $T_c$ of 69 K, for another
overdoped sample with $T_c$ = 60 K and in the inset data on
underdoped orthoII YBCO$_{6.50}$ with $T_c$ = 59 K. This material
is particularly well-ordered with every second chain full and the
others completely empty. In all cases two values of temperature
are shown, one in the superconducting state (dashed blue curve)
and the other in the normal state just above $T_c$ (solid red
curve). The difference between the behavior of underdoped and
overdoped samples is striking and can be understood\cite{hwang07}
as due to the opening of a pseudogap in the fully renormalized
density of state $\tilde{N}(\omega)$ of formula Eq. \ref{eq4}
which also determines the optical self-energy through equation Eq.
\ref{eq6} and Eq. \ref{eq7}. The prominent peak around 750
cm$^{-1}$ seen in both underdoped materials which is absent in the
overdoped case, can be traced to the opening of a gap in
$\tilde{N}(\omega)$ with lost states piled up in the energy region
just above $\omega=\Delta_{pg}$ as well as the existence of a
prominent boson mode in the electron-boson spectral density
$I^2\chi(\Omega)$. All three of the above conditions are needed.

%
%
\begin{figure}[t]
  \vspace*{-1.0 cm}%
  \centerline{\includegraphics[width=3.5 in]{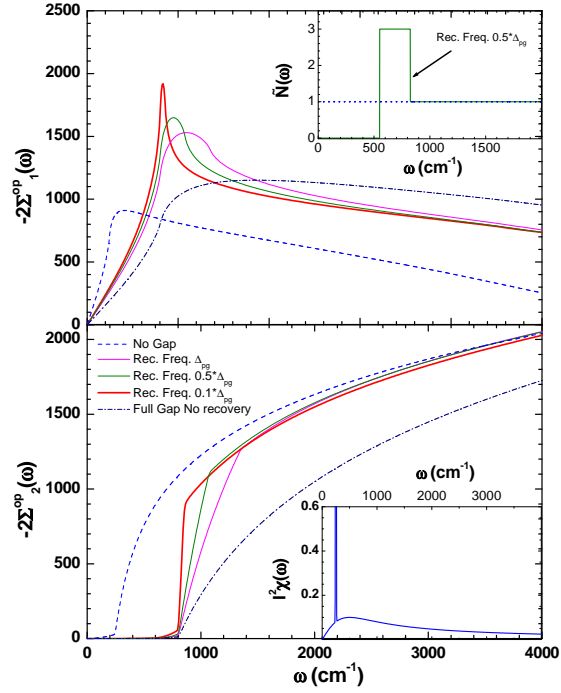}}%
  \vspace*{-1.0 cm}%
\caption{(color online) Model calculation of minus twice the real
and imaginary parts of the optical self-energy
$-2\Sigma^{op}_{1}(\omega)$ (top frame) and
$-2\Sigma^{op}_{2}(\omega)$ (bottom frame) in cm$^{-1}$ as a
function of $\omega$ also in cm$^{-1}$. All curves are based in
the electron-boson spectral density shown in the inset lower
frame. The dashed (blue) curve has no pseudogap while the
dash-dotted has a pseudogap $\Delta_{pg} =$ 550 cm$^{-1}$ but with
no recovery of states in the self consistent density of state
$\tilde{N}(\omega)$. The others have a recovery region right above
$\Delta_{pg}$ with conservation of total states applied. The
recovered states are piled up in the region $\Delta_{pg}$,
(1/2)$\Delta_{pg}$, and (1/10)$\Delta_{pg}$ for medium (purple),
light (olive), and heavy (red), respectively. The inset in the
upper frame shows the effective DOS $\tilde{N}(\omega)$ in the
case (1/2)$\Delta_{pg}$.}
 \label{Fig2}
\end{figure}

Fig. \ref{Fig2} shows results of model calculations for minus
twice the real part (top frame) and imaginary (bottom frame) part
of the optical self-energy ($\Sigma^{op}(\omega)$) for a system
with the electron-boson spectral density $I^2\chi(\omega)$ shown
in the inset of the lower panel. This model for the spectral
density (note the prominent resonance peak at 31 meV) is motivated
by an earlier study~\cite{hwang06} of orthoII YBCO$_{6.50}$ in
which $I^2\chi(\omega)$ was fit to its measured optical scattering
rate. What is different here is that we includes a pseudogap in
$\tilde{N}(\omega)$ as well as a recovery region just above
$\Delta_{pg}$ where the lost states in $\tilde{N}(\omega)$ are to
be found so as to conserve states in the DOS. The blue dashed
curve is for comparison and includes no pseudogap. The expressions
for $\lambda^{op}(\omega)$ and $1/\tau^{op}(\omega)$ in this
simple limit reduce to\cite{carbotte05}
\begin{equation}
\lambda^{op}(\omega) = \frac{2}{\omega}\int^{\infty}_{0} d\Omega
I^2\chi(\Omega)\Big{[}\frac{\Omega}{\omega}\ln\Big{|}\frac{\Omega^2-\omega^2}{\Omega^2}\Big{|}
+\ln\Big{|}\frac{\Omega+\omega}{\Omega-\omega}\Big{|}\Big{]}
\label{eq8}
\end{equation}
and\cite{knigavko06,allen71,shulga91}
\begin{equation}
\frac{1}{\tau^{op}(\omega)}=\frac{2
\pi}{\omega}\int^{\omega}_{0}d\Omega
(\omega-\Omega)I^2\chi(\Omega) \label{eq9}
\end{equation}

On comparing Eq. \ref{eq9} with the imaginary part in Eq.
\ref{eq3}, it is clear that for a simple delta function
$I^2\chi(\Omega)\equiv I_0\delta(\Omega-\Omega_E)$ which
represents coupling to a single Einstein mode at $\Omega_E$, the
optical scattering rate starts out of zero at $\omega = \Omega_E$,
then rises according to the factor $(\omega-\Omega_E)/\omega$ and
reaches its saturated value of $2 \pi I_0$ only for $\omega \gg
\Omega_E$. By contrast the quasiparticle scattering rate has a
discontinuous jump out of zero at $\omega = \Omega_E$ to its
saturated value and remains at this constant value for all
energies beyond this. This behavior for $1/\tau^{op}(\omega)$ is
seen in the (blue) dashed curve of Fig. \ref{Fig2} bottom frame.
Because we have also included a background in $I^2\chi(\omega)$ in
addition to a prominent peak at $\Omega_E =$ 250 cm$^{-1}$, there
are minor differences, including very small tails below the energy
of the prominent peak in $I^2\chi(\omega)$ shown in the inset. The
dot-dashed (blue) curve includes a pseudogap $\Delta_{pg} =$ 550
cm$^{-1}$ with the lost states in $\tilde{N}(\omega)$ moved to
infinity. In this case the main rise in $1/\tau^{op}(\omega)$ is
at $\Omega_E+\Delta_{pg} \simeq$ 800 cm$^{-1}$. Beyond this the
curve rises approximately like
$[\omega-(\Omega_E+\Delta_{pg})]/\omega$ which is less rapid than
the $(\omega-\Omega_E)/\omega$ curve for the $\Delta_{pg} = 0$
case.

The three remaining curves also have a pseudogap of 550 cm$^{-1}$
but, in addition the lost state in $\tilde{N}(\omega)$ are placed
in the energy region just above $\Delta_{pg}$ and state
conservation is respected. This causes the scattering rate
$1/\tau^{op}(\omega)$ to rise much faster than in the dot-dashed
curve because just above the gap we have more states to scatter
into. The steepness of the rise depends on the distribution of
states above $\omega = \Delta_{pg}$. The more the pileup is
restricted in range the steeper the rise. Medium (purple), light
(olive), and heavy (red) solid curves correspond, respectively to
the case when the missing states are placed between $\Delta_{pg}$
and $2\Delta_{pg}$, 1.5$\Delta_{pg}$, and 1.1$\Delta_{pg}$ (see
inset in the top frame for the case 1.5 $\Delta_{pg}$ where the
DOS is shown). Note also that the end of the recovery region in
all cases is mark with a kink in $1/\tau^{op}(\omega)$ after which
the remaining rise is much more gradual and smooth. The features
just described imply definite signatures in the corresponding real
part of the optical self-energy as these are related by
Kamers-Kronig (K-K) transform. The results for
$-2\Sigma^{op}_{1}(\omega)$ based on Eq. 6 are shown in the top
frame of Fig. \ref{Fig2}. As is known from the work of Carbotte,
Schachinger and Hwang\cite{carbotte05} the dashed (blue) curve
would have a logarithmic singularity in slope at $\omega =
\Omega_E$ if we were using a pure delta function model and would
have zero slope at $\omega = \sqrt{2}\Omega_E$. Similarly the
dash-dotted (blue) curve would have infinite slope at
$\Omega_E+\Delta_{pg}$ and zero slope at
$\sqrt{2}(\Omega_E+\Delta_{pg})$. These rules are very nearly
satisfied in our model calculation even though we are using the
spectra displayed in the inset lower frame rather than a pure
delta function. We recall that the K-K transform of a sharp step
like rise at $\Omega_E$ and constant after this as applies to the
quasiparticle scattering rate in a delta function model, has a
logarithmic singularity at $\omega = \Omega_E$. The heavy (red)
curve in the lower frame comes close to this case and indeed its
K-K transform shows a sharp peak at this frequency reminiscent of
a logarithmic singularity. We believe this to be the signature in
the real part of the optical self-energy of pseudogap
formation\cite{hwang07} as seen so prominently in the data of Fig.
\ref{Fig1} for the two underdoped samples. A model with recovered
DOS within $\Delta_{pg}$ shows a clearly identifiable hat type
structure in the medium solid (purple) curve for
$-2\Sigma^{op}_{1}(\omega)$ (Fig. \ref{Fig2} top frame) missing in
both (blue) curves. This hat is perhaps not quite as sharp in this
model calculation as it is in the data which is however less
peaked than the heavy (red) curve. This indicates that the
recovery region is consistent with a renormalized density of state
for which the conservation of states occurs on the scale of $\leq
\Delta_{pg}$.

Another interesting possibility to consider is the case when the
pseudogap does not reduce $\tilde{N}(\omega)$ to zero for $\omega
< \Delta_{pg}$ but rather still has a finite value. To illustrate
this possibility, in the top left hand frame of Fig. 3, we show
the imaginary part of $\Sigma^{op}(\omega)$ for the case when the
DOS is reduced to half its value rather than to zero below
$\Delta_{pg}$. The light solid (olive) curve is to be compared
with the heavy solid (red) curve which we reproduced from the
bottom frame of Fig. 2. Both curves include full recovery in the
energy interval $\Delta_{pg}$ and $2\Delta_{pg}$. The light solid
(olive) curve now starts at $\omega = \Omega_E$ and has a step at
$\Omega_E + \Delta_{pg}$ after which  it shows the characteristic
sharp rise which we have associated with the density of state
recovery region. In this case we see clear signatures of the
resonance mode and of the pseudogap recovery region separately.

\section{Effect of the pseudogap on real part of conductivity
and on its partial sum rule}

\begin{figure}[t]
  \vspace*{-1.0 cm}%
  \centerline{\includegraphics[width=3.5 in]{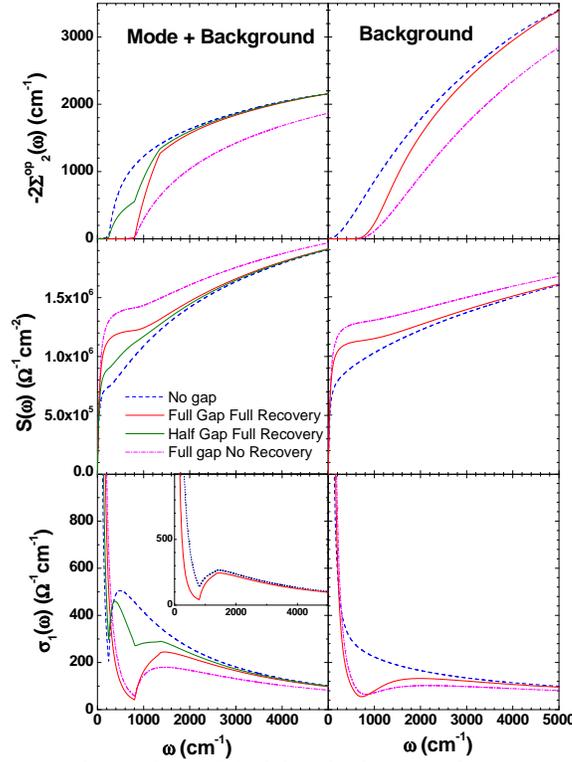}}%
  \vspace*{-1.0 cm}%
\caption{(color online) Model calculation of minus twice the
imaginary part of the optical self-energy
$-2\Sigma^{op}_{2}(\omega)$ (top frame), the partial optical sum
rule to energy $\omega$ in units of $\Omega^{-1}$cm$^{-2}$ (middle
frame) and the real part (absorptive) of the optical conductivity
in cm$^{-1}$ (bottom frame) as a function of $\omega$ in
cm$^{-1}$. The curves on the left were calculated with the
$I^2\chi(\omega)$ shown in the inset of Fig. 2 while the right
hand panel employed only the background contribution without a
sharp optical resonance mode. Dashed (blue) curve is for no gap,
dash-dotted (purple) for a gap without recovery, solid (red) for a
gap with recovery, and light solid (olive) for a model in which
only half the states are removed below the gap (with recovery).
Inset same as solid (red) curve but with 4 times more elastic
(impurity) scattering included.}
 \label{Fig3}
\end{figure}

Having a model for the optical self-energy $\Sigma^{op}(\omega)$,
we can calculate from it the real part of the conductivity. From
Eq. \ref{eq1} and the definitions of $1/\tau^{op}(\omega)$ and
$\lambda^{op}(\omega)$ we get
\begin{equation}
\sigma_1(\omega) = \frac{\Omega_p^2}{4
\pi}\frac{1/\tau^{op}(\omega)}{[\omega(1+\lambda^{op}(\omega))]^2+[1/\tau^{op}(\omega)]^2}
\label{eq10}
\end{equation}
One point that needs to be made is that at zero temperature the
inelastic optical scattering rate vanishes for $\omega < \Omega_E$
in an Einstein model and the optical conductivity Eq. \ref{eq10}
becomes pathological: $\sigma_1(\omega) =
\Omega_p^2\delta(\omega)/[4 (1+\lambda^{op}(0))]$ has a delta
function at $\omega = 0$ with weight reduced over the free
electron case by $1/(1+\lambda)$, where $\lambda \equiv
\lambda^{op}(0)$, which coincides with the quasiparticle mass
enhancement parameter $\lambda^{qp}(\omega = 0)$ even for the case
of an energy dependent density of state as we are considering
here. . This can be remedied by including a small amount of
elastic impurity scattering as we did in Eq. \ref{eq2}. In this
case $1/\tau^{op}(\omega)$ around $\omega = 0$ is a finite
constant and its corresponding real part vanishes. Results for
$\sigma_1(\omega)$ vs. $\omega$ are shown in the bottom frame of
Fig. \ref{Fig3}. In both cases $1/\tau^{imp}$ is set equal to 80
cm$^{-1}$ and $\Omega_p$ is 10000 cm$^{-1}$. For the left hand
frame we have used the $I^2\chi(\omega)$ shown in the inset of the
bottom frame of Fig. \ref{Fig2} which has an optical resonance as
well as a background while for the right hand frame only the
background is used. Starting with the left panel, the dashed
(blue) curve clearly shows two regions, the Drude plus a Holstein
boson assisted absorption piece which extends way beyond the Drude
and mirrors the spectral density $I^2\chi(\Omega)$. For a delta
function its onset is at $\omega = \Omega_E$ and grows out of zero
according to Eq. \ref{eq10}. It contains $\lambda/(1+\lambda)$ of
the optical spectral weight. The remaining optical spectral weight
$1/(1+\lambda)$ is to be found in the coherent Drude contribution.
The width of the Drude is given very nearly by
$[1/\tau^{imp}(1+\lambda)]$ which is different for the various
curves because $\lambda$ varies as we will explain. In the inset
(bottom left frame) we repeat the solid (red) curve for pseudogap
plus recovery and compare it with the same case but now the
elastic scattering has been increased by a factor of 4. This fills
in the region between Drude and Holstein processes much as is seen
in the Bi-2212 data as one goes further into the underdoped region
to $T_c =$ 69 K\cite{hwang07a}. This can effectively switch some
of the extra spectral weight that was transferred to the coherent
part of the conductivity (Drude part) by the opening of the
pseudogap back to the larger energies associated with the
incoherent boson assisted part of the absorption (Holstein)
region.
%
%

When there is no pseudogap as in the dashed (blue) curve, the
quasiparticle mass enhancement parameter is given by
\begin{equation}
\lambda = 2\int^{\infty}_{0}d\Omega\frac{I^2\chi(\Omega)}{\Omega}
\label{eq11}
\end{equation}
If however a pseudogap is introduced it becomes modified and reads
instead from Eq. \ref{eq5} in the limit $\omega \rightarrow 0$
\begin{equation}
\lambda = 2
\int^{\infty}_{0}d\omega'\tilde{N}(\omega')\int^{\infty}_{0}
d\Omega\frac{I^2\chi(\Omega)}{(\omega'+\Omega)^2} \label{eq12}
\end{equation}
To remain simple we assume $\tilde{N}(\omega) = t$ for $\omega <
\Delta_{pg}$ and $2-t$ for $\omega \in (\Delta_{pg},
2\Delta_{pg})$ and 1 beyond. That is, we have piled up the missing
state equally in the interval $\Delta_{pg}$ to $2\Delta_{pg}$. In
this case the mass enhancement parameter becomes
\begin{equation}
\lambda = 2\int^{\infty}_{0}d\Omega\frac{I^2\chi(\Omega)}{\Omega}
h(\Omega) \label{eq13}
\end{equation}
with the modulating factor $h(\Omega)$ equal to
\begin{equation}
h(\Omega)=\frac{2t+3\Omega/\Delta_{pg}+(\Omega/\Delta_{pg})^2}{(1+\Omega/\Delta_{pg})(2+\Omega/\Delta_{pg})}
\label{eq14}
\end{equation}
for the fully recovered case and
\begin{equation}
h(\Omega)=\frac{t +\Omega/\Delta_{pg}}{1+\Omega/\Delta_{pg}}
\label{eq15}
\end{equation}
for the case of no recovery region. This factor in effect reduces
the low $\Omega$ contribution to the mass enhancement factor from
the spectral density alone. For a full pseudogap the suppression
provides an extra factor of $3\Omega/(2\Delta_{pg})$ and
$\Omega/\Delta_{pg}$ at small $\Omega$ while for finite $t$ it
gives $t+3\Omega/(2\Delta_{pg})$ and $t+\Omega/\Delta_{pg}$,
respectively. For the specific case considered here $\lambda =$
2.37, when we include a pseudogap Eq. \ref{eq12} is reduced to
0.78 for a full gap with no recovery region above it and it is
1.06 when recovery is included as in Eq. \ref{eq14}. An unexpected
consequence of this reduction in $\lambda$ is that the Drude
contribution (coherent part) to $\sigma_1(\omega)$ is increased
when a pseudogap is included and there is a corresponding decrease
in the boson assisted Holstein (incoherent) contribution centered
off $\omega = 0$. These factors translate into a wider Drude as
$\lambda$ is decreased. Dash-dotted (purple) is widest, then solid
(red) and finally dashed (blue) at the same time the Holstein
boson assisted region shows increasing weight from dash-dotted
(purple) to solid (red) to dashed (blue) curves. Note also that a
complete pseudogap below $\omega=\Delta_{pg}$ cuts off the
Holstein region which now starts at $\omega =
\Delta_{pg}+\Omega_E$ in the both solid (red) and dash-dotted
(purple) curves. The light solid (olive) curve is different from
the others and shows two peaks rather than one in the Holstein
region. It corresponds to an incomplete pseudogap with height $t
=$ 0.5 at $\omega = 0$. This translates into a larger $\lambda$
than for the solid (red) curve and its Drude peak indeed falls
between (red) solid and (blue) dashed curves. Its Holstein region
however starts at $\Omega_E\simeq 250$ cm$^{-1}$ (position of
large peak in $I^2\chi(\omega)$ of the inset in lower frame of
Fig. \ref{Fig2}) because the DOS is finite at the Fermi surface.
This onset is followed by a peak with a second onset seen clearly
at $\omega = \Delta_{pg}+\Omega_E$ which corresponds to the sudden
increase in the electronic DOS. To end this discussion of the
effect of a pseudogap on the real (absorptive) part of the optical
conductivity we consider the lower right hand panel of Fig. 3.
Here only the background spectrum in $I^2\chi(\omega)$ is included
{\it i.e.} the delta function like optical resonance is excluded.
We see that now Drude and Holstein contribution in the dashed
(blue) curve are not clearly separated. While a two-contribution
structure is seen in the other two curves these are not as well
defined as in the corresponding curves of the left-hand panel. The
existence of a sharp peak in electron-boson spectral density helps
separate out the two distinct absorption processes (Drude and
Holstein).

In the middle frame of Fig. 3 we show results for the partial
optical sum $S(\omega)$ defined as
\begin{equation}
S(\omega) = \int^{\omega}_{0}d\omega'\sigma_1(\omega').
\label{eq16}
\end{equation}
For $\omega \rightarrow \infty$ in Eq. \ref{eq16}, we get the
usual complete sum rule with $S(\omega \rightarrow \infty) =
\Omega_p^2/8$. In our units this corresponds to $\simeq
2.6\times10^6$ cm$^{-2}$ with $\Omega_p =$ 10000 cm$^{-1}$. One
sees clearly in these curves a rapidly increasing Drude
contribution followed by a flattened region and then the Holstein
contribution setting in at large values of $\omega$. The fraction
of spectral weight seen in the Drude region is $1/(1+\lambda)$ of
the total contribution. The remainder $\lambda/(1+\lambda)$ is the
boson assisted contribution. If these two regions were truly
separated in $\sigma_1(\omega)$ the flattened region noted above
would be perfectly flat. We make one more point about these
results. In all cases there is, of course, conservation of optical
spectral weight but this conservation occurs over a much larger
frequency region for the real part of the conductivity than for
the density of states itself which, in our model, is limited to
the range 0 to $2\Delta_{pg}$. This is also true of the imaginary
part of the optical self-energy but not for $\sigma_1(\omega)$ or
the partial sum $S(\omega)$ of Eq. \ref{eq15}. Yu {\it et
al.}\cite{Yu07} have recently noted this in their c-axis optical
study of spectral weight redistribution due to the pseudogap.

Finally returning to the top frame of Fig. 3 we note that the
sharp onset in scattering rate seen in the left-hand panel is
considerably smeared out when the resonant peak in the boson
spectral density $I^2\chi(\omega)$ shown in the inset of the lower
frame of Fig. 2 is left out and only the background spectral
density is employed. This is shown in the right-hand panel
(compare color coded curves).

\section{Conclusions}

Motivated by the observation of hat like peak structures seen in
the real part of the optical self-energy in underdoped cuprates
which are absent in the overdoped case, we have considered a
pseudogap model for their electronic structure consisting of a
simple gap ($\Delta_{pg}$) in the electronic DOS about the Fermi
energy, with missing states recovered in the energy region right
above it. This simple model augmented with an optical peak in the
electron-boson spectral density is remarkably successful in
describing the data and is further used to describe other optical
quantities. The optical scattering rate, the real part of the
conductivity and its partial sum are considered as is the
derivative of $\Sigma^{op}_1(\omega)$. These quantities all show
specific signatures of the pseudogap. While the redistribution of
electronic states in the DOS is limited to the region $\omega \leq
2\Delta_{pg}$ we found that the corresponding redistribution of
optical spectral weight in the real part of the conductivity is
spread over a much large range. This is also true for the real
part of the optical self-energy but not for its imaginary part for
which the important changes due directly to the opening of
pseudogap are confined much more to the range $\omega \leq
2\Delta_{pg}$. Comparing results of model calculations we conclude
that pseudogap effects can be readily distinguished from the
effect of multiple peaks in the electron-boson spectral density
$I^2\chi(\omega)$.

\ack

This work has been supported by the Natural Science and
Engineering Research Council of Canada and the Canadian Institute
for Advanced Research.

\end{document}